# A simple configuration setup for compton suppression spectroscopy

N. X. Hai[1,] N. N. Dien[1], P. D. Khang[2], V. H. Tan[2], N. D. Hoa[3]

[1]*Nuclear Research Institute, 01 Nguyen Tu Luc, Dalat, Vietnam*

[2]*Nuclear Training Center, 140 Nguyen Tuan, Hanoi, Vietnam*

[3]*University of Dalat, 01 Phu Dong Thien Vuong, Dalat, Vietnam*

**Abstract:** The fast timing, standard timing and easy timing are popular timing configurations of compton suppression spectroscopy. Such spectroscopes always use a module of coincidence or time-to-amplitude converter (TAC). A compton suppression spectroscopy with "semi-timing" configuration is presented in this paper. The "semi-timing" configuration is relatively simple and easy system setup, especially this spectroscopy does not need to use module of coincidence or TAC. The performance of spectroscopy was tested and summarized. The count rate background, full peak efficiency and the ratios of area/background of peaks in suppressed and unsuppressed modes were comparative.

**I. Introduction**

In gamma-ray spectrometry, compton suppression is a technique that reduces the background by preventing registration of signals from gamma-ray, scattered from the main detector. This technique is not new as the construction of spectrometer employing active shielding of the main detector for the suppression of compton continuum started in early times of the 1950s for research in nuclear physics as well as for low-level counting in environmental studies [1]. The main advantage of compton suppression system is shielding radiation from the sample which scattered in the main detector. At early beginnings, the main detectors were NaI(Tl) scintillator. Later on, with the appearance of the semiconductor detectors with their high energy resolution capability, the main detectors have become of high purity germanium (HPGe) material. The various design configurations and the historical development of the compton suppression systems have been reviewed in the some studies [1].

Today, many electronic configurations of compton suppression spectroscopy have been developed, all of them are based on a module of coincidence. The applications and compton background reducing-capacity of compton suppression spectroscopy were mentioned in Refs.

---

[1] *Corresponding author: Email address: nxhai@hcm.vnn.vn*

[1-6]. Due to the lager number of gamma rays incident on the main detector, the associated compton continuum is a significant hindrance for low background. For example, the compton continuum makes the search of low-intensity peaks difficult and increases the uncertainty of the measured activities. The reduction capacity of compton background depends on detector type, electronic configuration, and timing method. In this study, a compton suppression spectroscopy has been set-up and installed at 500 kW Research Reactor in Dalat for neutron activation analysis and nuclear data measurement. The reduction of the compton continuum has been achieved by surrounding the HPGe detector with the Bismuth Germanate (BGO) detectors (guard detectors) whose signals are used for the anti-coincidence gating in the analog-to-digital converter (ADC). The timing circuit was "semi-timing" which is used for timing of BGO detectors. In this configuration, the modules of coincidence and TAC weren't used, so it is less expensive and easier adjustment than fast timing configuration.

## 2. The compton suppression spectrometer setup

*2.1. Instrument and configuration*

The data acquisition part of the compton suppression system adds more complication beside the requirements for the specific detector geometry in order to achieve improved spectra information. The anti-coincidence instrumental arrangement between the two detectors requires the timing analysis of the signals. The fast timing measurements are desired in such timing analysis for accurate action. With fast timing measurements, an additional circuit (beside the ordinary energy circuit) is used to study the time correlation between the two signals from both detectors.

Like in any system dedicated for timing measurements, the compton suppression technique relies on the timing properties of the employed detectors and the processing electronics used for the anti-coincidence rejection of pulses resulting from the same event which are detected by both detectors simultaneously. The radiation measurement systems with fast timing response are always desired in studies involving timing measurements. The fast coincidence circuits in which the signal from central detector is separated in two branches, one is a slow branch for energy analysis and the other is a fast branch for the timing measurement, are used for this purpose. Ordinary preamplifiers and amplifiers are used in the energy branch whereas fast timing filter amplifiers and timing discriminators are required for timing analysis in the timing branch [1-6]. There are some types of simple instrumental arrangements of less complex chains, which may provide equivalent results without the need for special setup procedures or adjustments [5, 6]. The compton suppression spectroscopy does not require fast timing and fast coincidence electronics in order to provide good suppression, coincidence resolving times of 0.5 to 1.0 μs are more than adequate considering HPGe detectors that create typically with amplifier time constants of 2-4 μs with resultant pulse widths (and dead times) of 8-20 μs. Nevertheless, some predictable and consistent coincidence (anti-coincidence) electronics set-ups are required to ensure good performance. Often experiments require fast timing and good timing resolution for other reasons and in such cases the timing is available for use in compton suppression [6].

The fast timing, standard timing and easy timing are popular types of timing. They have been developed and used for various applications [1-6]. In this study, a new timing configuration for compton suppression spectroscopy, which is called "fast semi-timing" configuration, is presented. The configuration of the spectroscopy is shown in Fig. 1.

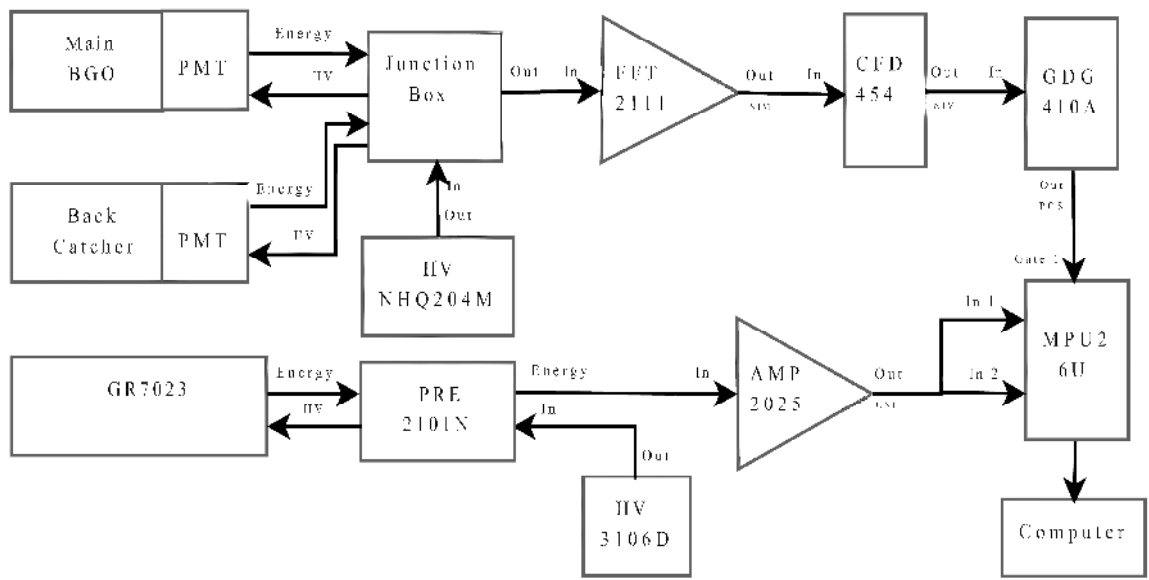

Fig. 1. Configuration of the compton suppression system.

*The central detector:* The central detector is a GR7023 Canberra n-type coaxial high-purity germanium detector (HPGe). It has an energy resolution (FWHM) of 2.36 keV and a peak-to-compton ratio of 58:1 for the 1.33 MeV $^{60}$Co peak. The relative efficiency is 72%.

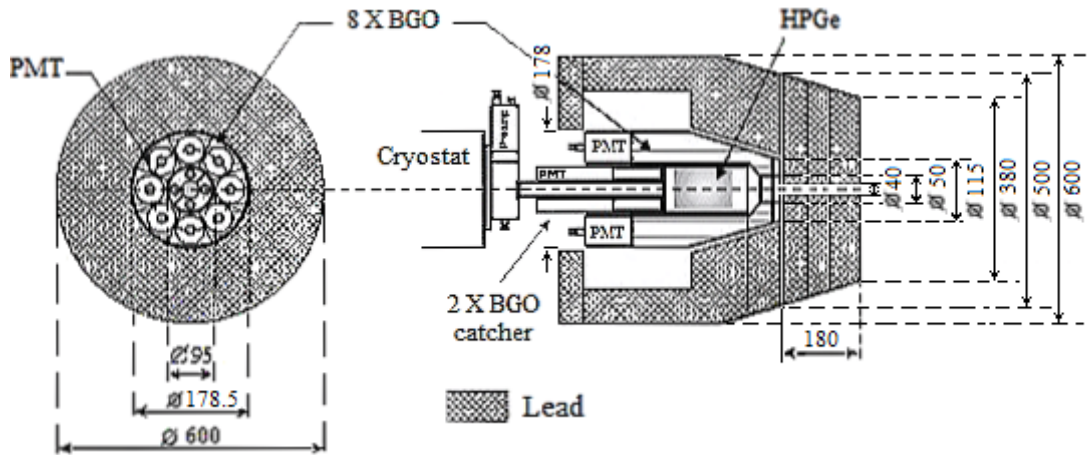

Fig. 2. The back and the cross-sectional view of the detectors and shielding system.

*The guard detectors:* The guard detectors are BGO, configured as Fig. 2. The main BGO detectors are within an aluminum housing with 0.8 mm thickness. The diameter of the hole at the front of the detector and the diameter of the detector endcap are 50 mm and 95 mm respectively. The length, the thicknesses of the side and the conical front are 200 mm, 27.5 mm and 20 mm respectively. It consists 8 crystal segments, each segment has read-out with photon multiplier tube (PMT) of Hammatsu type R1924. The energy resolution is less than 17% at 662 keV $^{137}$Cs peak.

The back-catcher BGO detectors are divided into 2 halfs, each half has read-out with 2 one-inch diameter PMT of Hammatsu type R1924. The inner diameter, the outer diameter and thickness of crystal are 86 mm, 95 mm and 50 mm respectively.

*Lead shield for the detector system:*

All parts of the detectors including the photomultiplier tubes are shielded by lead of 10 cm thickness at least. A lead-stepped collimator is located in the front of the opening of the guard detectors. The length and inner diameter of the lead collimator are 180 mm and 40 mm respectively.

*The electronic modules:*

The electronic modules are manufactured by Canberra except the high voltage module for BGO detectors, which were produced by Fast Comptec. They include 2111 fast filter timing amplifier (FFT), 454 quad constant fraction discriminator (CFD), 2025 main amplifier (AMP), 3106D high voltage power supply, 410A gate delay generator (GDG), 2101N preamplifier, MPU2-6U multiport multichannel analyzer, NHQ204M high voltage power supply, and junction box for joining the signals from PMT's and connecting them to NIM modules.

2.2. *Compton suppression spectroscopy setup*

The Fig. 1 presents the configuration of the compton suppression spectroscopy with fast semi-timing.

*The direct signal*

The energy output signal of 2101N preamplifier is connected to normal input of 2025 AMP. The parameters of 2025 AMP are set at 6 μs, positive input and around 0.5 keV per channel. The unipolar output of 2025 AMP is connected to the first and second input of MPU2-6 (the first channel for compton suppression mode and second channel for single mode). The MPU2-6 is connected to computer through USB port. The parameters of ADCs are set at 16k and anti mode. The Genie 2000 control software is used.

*The gate signal and adjustments*

The signals from the guard detectors are fed via the FFT amplifier to the CFD. The shaping time constants of the FFT amplifier are set at values of differentiate equal 50 and integrate equal 50. The gain of the FFT amplifier is set near maximum in order to enable the discrimination at a level just above the noise. The output of the CFD is connected to the GDG, which opens the gate of the ADC.

## 3. Testing, results and discussion

*3.1. Testing*

In order to characterize the performance of the compton suppression spectroscopy with fast semi-timing configuration, the gamma-ray background from shielding materials and environment and the radioisotope sources were measured in two modes: with and without compton suppression. The Single/Suppression measurements were made under the same conditions, measurement geometry, measurement time and time constant. During experiment, all isotope sources were located outside the guard detectors and collimated to prevent direct interaction with the guard detectors.

The $^{60}$Co and $^{152}$Eu sources were used to evaluate the peak-to-compton ratio (P/C) and compton suppression factor (SF) and total peak efficiency ratio (TPE).

The P/C is defined as the ratio of the counts/channel in the highest full-energy peak to the counts/channel in a typical channel of the compton continuum associated with the full-energy peak. With $^{60}$Co spectra, the compton region used for P/C calculations has been defined in IEEE standard 325 for $^{60}$Co as 1040 keV to 1096 keV. Therefore the formula for P/C is as follows [7]:

$$P/C = \frac{\text{Number of counts in highest channel of 1.33 MeV peak}}{\text{Average counts per channel (1040 keV and 1096 keV)}} \qquad (1)$$

The SF is defined as counts/channel of the single mode divided by counts/channel of the compton suppression mode.

The TPE is defined as peak count rates of the single mode divided by peak count rates of the compton suppression mode. Furthermore, the ratio of gross peak with and without compton suppression has been calculated, that evaluated the effect of continuum background on area of full-energy peaks.

*3.2. Results*

*Background of shielding materials and environment with/without suppression:*

The suppressed/unsuppressed background spectra were collected for 65,000 seconds.

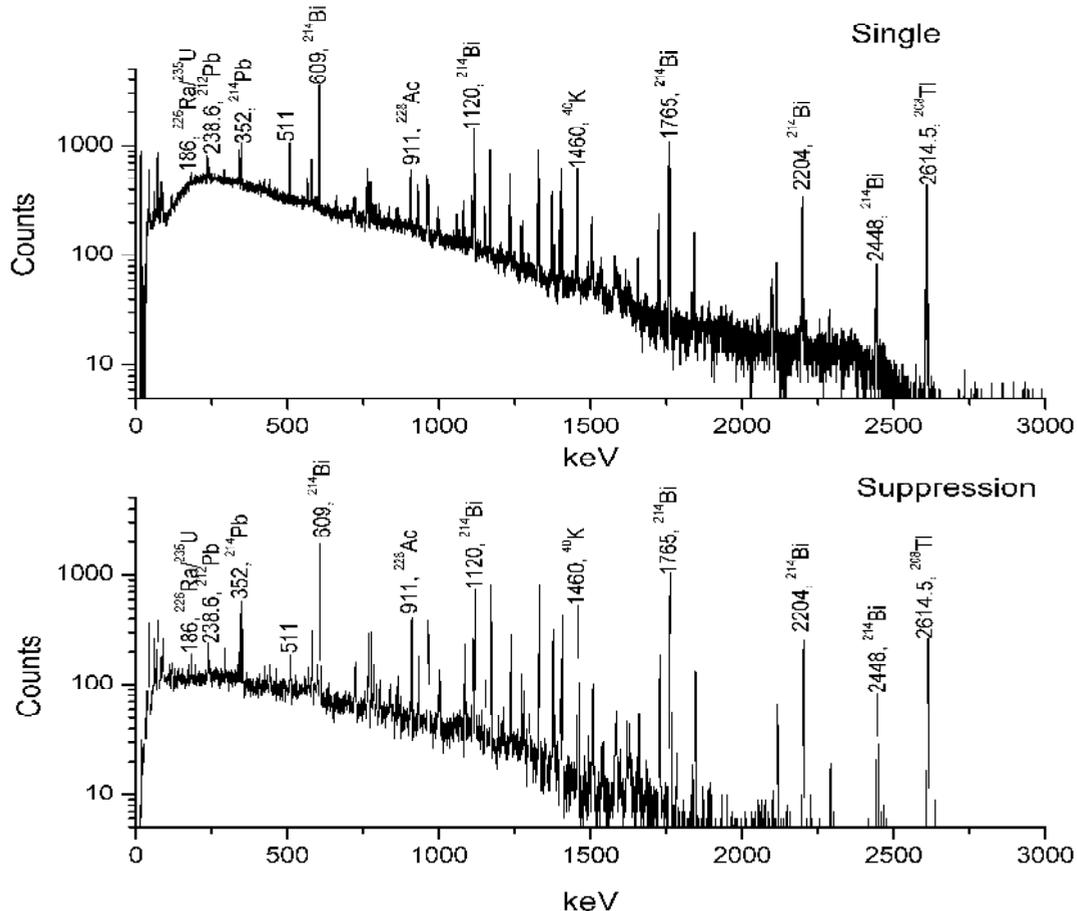

Fig. 3. Single and suppression of background spectra.

With the unsuppressed/suppressed spectra, the integral count rates for the energy region of 15-3000 keV were 12.84 count per second (cps) and 4.26 cps respectively. An overall reduction of 66.82 % has been achieved as an improvement in the background reduction within this energy range. Peak area of the main peaks observed in Fig. 3 is listed in Table 1. The S501C of Genie 2000 software was used to calculate. The statistical uncertainties were quote and same peak is evaluated on background different for a factor about 3. The difference of count rates of some peaks in two cases were outside the scope of statistical uncertainties.

Table 1. Peak area of main peaks in natural background spectrum which measured with single and compton suppression mode.

| Energy (keV) | Nuclide | Peak area (cps) | | Ratio of Single/Suppression |
|---|---|---|---|---|
| | | Single | Suppression | |
| 185.7 | $^{226}$Ra/$^{235}$U | 0.0097 ± 4.0% | 0.0051 ± 5.5% | 1.89 ± 6.8% |
| 238.6 | $^{212}$Pb | 0.0147 ± 3.3% | 0.0083 ± 4.3% | 1.77 ± 5.4% |
| 351.9 | $^{214}$Pb | 0.0381 ± 2.0% | 0.0308 ± 2.2% | 1.24 ± 3.0% |
| 511 | Annihi./$^{208}$Tl | 0.0761 ± 1.4% | 0.0099 ± 4.0% | 7.67 ± 4.2% |
| 569.7 | $^{207M}$Pb | 0.0130 ± 3.5% | 0.0032 ± 6.9% | 4.01 ± 7.7% |
| 583.1 | $^{208}$Tl | 0.0299 ± 2.3% | 0.0148 ± 3.2% | 2.02 ± 4.0% |
| 609.3 | $^{214}$Bi | 0.2117 ± 0.9% | 0.1219 ± 1.1% | 1.74 ± 1.4% |
| 911.1 | $^{228}$Ac | 0.0330 ± 2.2% | 0.0298 ± 2.3% | 1.11 ± 3.2% |
| 1063.6 | $^{207M}$Pb | 0.0056 ± 5.3% | Not detected | - |
| 1120.3 | $^{214}$Bi | 0.1038 ± 1.2% | 0.0587 ± 1.6% | 1.77 ± 2.0% |
| 1460.8 | $^{40}$K | 0.0495 ± 1.8% | 0.0492 ± 1.8% | 1.01 ± 2.5% |
| 2614.5 | $^{208}$Tl | 0.0516 ± 1.7% | 0.0341 ± 2.1% | 1.51 ± 2.8% |

The results of Table 1 can be summarized as follow:

The ratios of single/suppression at 185.7, 238.6 and 351.9 keV are significantly higher than unity, although they do not belong to cascades. This showed the poor gating at low energies, the small amplitude of PMT's output and did not use preamplifier could be a reason.

In spite of full-energy peaks formed by gamma-ray of isotopes such as $^{214}$Pb, $^{214}$Bi they are belong to cascades, the ratio significantly higher than unity could be supporting the point of emission close to the detectors.

Contrary to above, the full-energy peaks of $^{207M}$Pb are strongly suppressed with compton suppression mode, because of that $^{207M}$Pb is contaminant of the BGO crystal and the total detection efficiency of the BGO system is very high.

The ratio of peak area of the doublet of the 511 keV is 7.67, that is considerably larger compared to 1.51 of the 2614.5 keV gamma-rays from $^{208}$Tl. This results shows that the peak area of 511 keV with single mode is formed mostly by annihilation radiation generated in/or near the BGO detector. The 511 keV peak and the single/double escape peaks corresponding to the 2614 keV from $^{208}$Tl have not been observed at all by the measurement with $^{228}$Th source in the compton suppression mode. Because of that the detection efficiency of the BGO detector system surrounding the HPGe with almost $4\pi$ is very high for annihilation radiation.

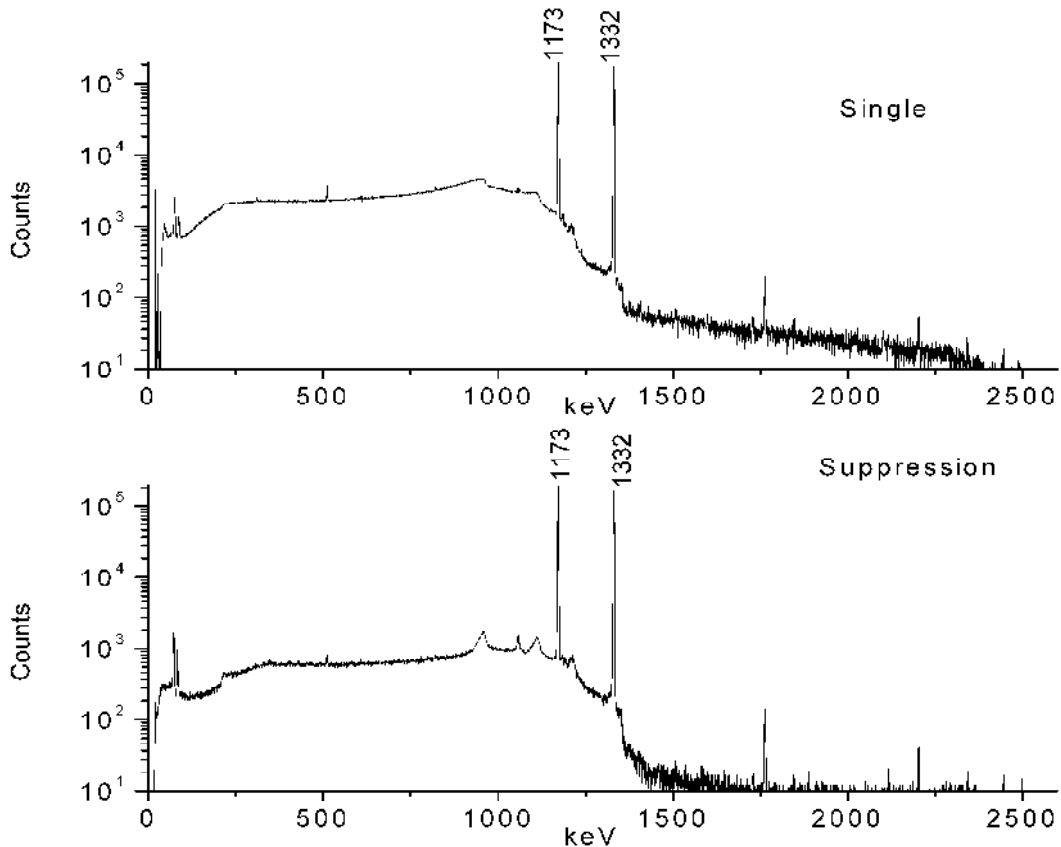

Fig. 4. Single and suppression of $^{60}$Co spectra.

The Fig. 4 represents the single/suppression of $^{60}$Co spectra. In this example, with single and compton suppression mode, the integral count rates for overall spectra were 788 cps and 368 cps respectively. The differences of the counts in the full energy peaks are 1.26 % (1173 keV) and 0.23 % (1332 keV) while the compton background (integral count from 15 keV to 1130 keV) is reduced by a factor of about 72%. At the lowest point in the compton continuum the SF is 5.7 and it is 5.2 in near the backscatter peak. The single/suppression P/C is 3.25 (152/48).

Furthermore, the Fig. 4 shows the relationship between the scattering angle and the resultant compton continuum in HPGe detector spectrum. The spectrum region below the channel 500 is distorted, because of that scattered gamma-ray has lost little energy in the HPGe detector and are traveling forward with high residual energy. The compton continuum in this region is depended on the total efficiency of the back-catcher BGO detectors. In the spectrum region from 1050 to 1200 keV, the scattered gamma-rays have lost most of their energy in the HPGe detector and are scattered backward. In this direction, the scattered backward gamma-ray could be out of detectors that rise compton edges. Some addition small peaks occur above the compton edge, they could belong to multiple scattering events, which are not detected by guard detectors.

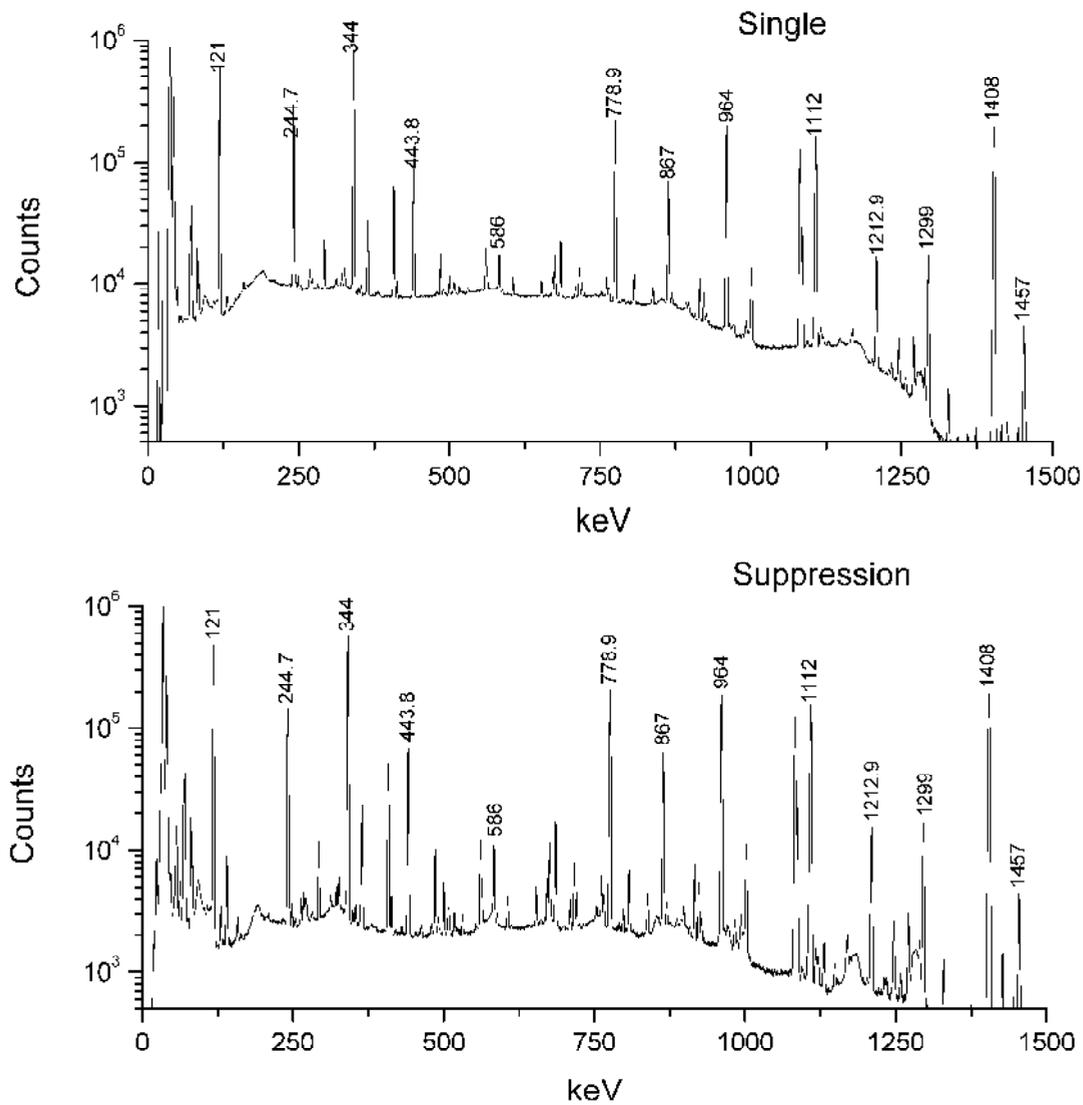

Fig. 5. Single and suppression of $^{152}$Eu spectra.

The Fig. 5 presents Single/Suppression of $^{152}$Eu spectra. The integral count rates for overall spectra were 349 cps and 535 cps respectively. At the lowest point in the compton continuum and the near the backscatter peak, the SF is about 4 and 3.5 respectively. The TPE values and ratio of gross area with Single/Suppression compton are shown in Table 2.

Table 2. The Single/Suppression ratio of peak area and gross area measured with $^{152}$Eu.

| Energy (keV) | TPE | Single/Suppression gross area |
| --- | --- | --- |
| 121.78 | 1.13 ± 0.10% | 1.16 ± 0.10% |
| 244.7 | 1.75 ± 0.17% | 1.92 ± 0.16% |
| 295.93 | 1.45 ± 0.68% | 1.84 ± 0.39% |
| 344.28 | 1.30 ± 0.09% | 1.35 ± 0.08% |
| 367.78 | 0.96 ± 0.45% | 1.50 ± 0.32% |
| 411.15 | 1.15 ± 0.29% | 1.63 ± 0.24% |
| 443.88 | 1.11 ± 0.26% | 1.52 ± 0.22% |
| 488.66 | 1.18 ± 0.76% | 2.48 ± 0.48% |
| 503.39 | 1.06 ± 1.22% | 2.85 ± 0.56% |
| 586.29 | 0.96 ± 0.73% | 1.90 ± 0.41% |
| 674.67+678.58 | 1.03 ± 0.62% | 1.92 ± 0.36% |
| 688.67 | 1.00 ± 0.54% | 1.63 ± 0.35% |
| 778.9 | 1.01 ± 0.15% | 1.13 ± 0.14% |
| 810.45 | 0.94 ± 0.96% | 2.48 ± 0.50% |
| 867.38 | 1.01 ± 0.26% | 1.25 ± 0.23% |
| 919.4 | 0.99 ± 0.85% | 1.66 ± 0.51% |
| 964.13 | 1.00 ± 0.15% | 1.05 ± 0.14% |
| 1005.28 | 1.00 ± 0.64% | 1.41 ± 0.43% |
| 1085.91+1089.70 | 1.01 ± 0.17% | 1.12 ± 0.16% |
| 1112.12 | 1.00 ± 0.16% | 1.07 ± 0.15% |
| 1212.95 | 1.01 ± 0.50% | 1.27 ± 0.43% |
| 1249.95 | 0.98 ± 1.36% | 1.74 ± 0.79% |
| 1299.12 | 0.96 ± 0.49% | 1.06 ± 0.44% |
| 1408.01 | 1.01 ± 0.14% | 1.01 ± 0.14% |
| 1457.63 | 1.01 ± 0.85% | 1.05 ± 0.78% |
| 1528.12 | 1.02 ± 1.17% | 1.11 ± 1.10% |

*3.3. Discussion*

With the suppressed compton configuration, the total background count rate which came from shield materials and environment was reduced about three times. Based on the spectra in the Fig. 4 and Fig. 5, the peak-to-compton ratio increased more than three times. The efficiency of full energy peaks in suppressed compton mode was a function of energy.

Since a collimated beam and the source are far from the detector, the peak count rates should be the same in the suppressed and unsuppressed spectrum. Therefore the TPE ratio should be around 1. The argument that the distortion of the background affects the TPE ratio at low energies is not sufficient. The background is more than an order of magnitude smaller than the peaks. Therefore its shape can't influence the peak areas for more than 10%. However the background spectrum of compton suppression spectroscopy is depended on purities of materials which are used to make detectors, shields and guard detectors.

The TPE ratio of lines from impurities is large if they belong to cascade decays, since they are emitted close to the detectors. The TPE ratio at low energies is large because the gating at low energies is poor.

## 4. Conclusion

This study describes a simple configuration of compton suppression spectroscopy. It is found that the compton suppression spectroscopy can be operated well without the module of coincidence or TAC. This timing configuration is quite useful for applications of neutron activation analysis and fundamental research. Furthermore, it can be good for low-level measurement. The reduction of guard detector's preamplifier could be distorted background at low energies.

Especially, the fast semi-timing configuration is suitable for compton suppression spectroscopy with fast preamplifiers which haven't timing output.

**Acknowledgements**

The authors thank the International Atomic Energy Agency and the Vietnam Ministry of Science and Technology for their support in the framework of the TC project RER/4/028 - VIE/9/012.